\documentclass[aps,prd,onecolumn,superscriptaddress]{revtex4}
\usepackage{amsfonts,amssymb,amsmath,graphicx,bm,hyperref}

\begin{document}
\title{Proton spin in a light-cone chiral quark model}

\newcommand*{\PKU}{School of Physics and State Key Laboratory of Nuclear Physics and Technology, Peking University, Beijing 100871, China}\affiliation{\PKU}
\newcommand*{\SBU}{Department of Physics and Astronomy, Stony Brook University, Stony Brook, NY 11794-3800, USA}\affiliation{\SBU}
\newcommand*{\CHEP}{Center for High Energy Physics, Peking University, Beijing 100871, China}\affiliation{\CHEP}

\author{Xinyu~Zhang}\affiliation{\PKU}\affiliation{\SBU}
\author{Bo-Qiang~Ma\footnote{Corresponding author. Email address: \texttt{mabq@pku.edu.cn}}}\affiliation{\PKU}\affiliation{\CHEP}%\affiliation{\CHPS}

\begin{abstract}
We discuss the spin structure of the proton in a light-cone
treatment of the chiral quark model. Based on the fact that the
quark helicity ($\Delta q$) measured in polarized deep inelastic
scattering experiments is actually the quark spin defined in the
light-cone formalism rather than the quark spin ($\Delta q_{QM}$)
defined in the conventional quark model (or in the rest frame of
the nucleon), we calculate the $x$ dependence of the polarized quark
distribution functions $\Delta q(x)$, and the polarized structure
functions $g_1 (x)$. Special attention is focused on the Melosh-Wigner rotation due to
the transversal motions of quarks inside the nucleon and Melosh-Wigner rotation effects
on the bare quark input. It is shown that our results match the
experimental data well.
\end{abstract}
%13.88.+e: Polarization in interactions and scattering, 12.39.Fe: Chiral Lagrangians, 14.20.Dh: Protons and neutrons

\maketitle

\section{Introduction}

It is well known that the nucleon consists of quarks and gluons, and the underlying theory describing their interaction is quantum chromodynamics (QCD). However, due to the nonlinear and nonperturbative nature of QCD, it is still beyond our practical abilities to calculate from the QCD Lagrangian directly to study the low-energy properties of the nucleon at present, and consequently there are many problems that remain unsolved. Among these problems, the spin structure of the nucleon is a particularly important one and has triggered a vast number of theoretical and experimental investigations.

In the SU(6) quark model, the proton wave function is given by
\begin{eqnarray}
|p^{\uparrow} \rangle &=& \frac{1}{\sqrt{2}}|u^{\uparrow} (ud)_{S=0}\rangle + \frac{1}{\sqrt{18}}|u^{\uparrow} (ud)_{S=1}\rangle - \frac{1}{3}|u^{\downarrow} (ud)_{S=1}\rangle \nonumber\\ &-& \frac{1}{3}|d^{\uparrow} (uu)_{S=1}\rangle + \frac{\sqrt{2}}{3}|d^{\downarrow} (uu)_{S=1}\rangle.
\label{wavefunction}
\end{eqnarray}
One can count the number of quarks with certain helicity, $u^{\uparrow} = 5/3$, $u^{\downarrow} = -1/3$, $d^{\uparrow} = 1/3$, $d^{\downarrow} = -2/3$, and $s^{\uparrow} = s^{\downarrow} = 0$. Naively, one may expect that the sum of quark helicity to be
\begin{eqnarray}
\frac{1}{2} \Delta \Sigma &=& \frac{1}{2}\left[\Delta u + \Delta d + \Delta s \right] \nonumber\\
&=& \frac{1}{2}\left[\left(\frac{5}{3}-\frac{1}{3}\right) + \left(\frac{1}{3}-\frac{2}{3}\right) + 0 \right] \nonumber\\
&=& \frac{1}{2},
\end{eqnarray}
which is equal to the proton spin. However, it was observed in the polarized deep inelastic scattering (pDIS) experiment by the European Muon Collaboration~\cite{Ashman:1987hv} that the sum of quark helicity was much smaller than $1/2$. This is the so-called ``proton spin crisis" or ``proton spin puzzle." Although many models have been proposed to explain this phenomenon, the issue is still far from being settled. In particular, we still need theoretical calculation of the $x$-dependent polarized quark distribution functions, with the expectation that these will match the experimental data well, since in most previous papers only integrated quantities were discussed, or only parametrizations were given by fitting the data.

As emphasized by one of us~\cite{Ma:1991xq,Ma:1992sj}, the quark spin is not Lorentz invariant, and the quark helicity ($\Delta q$) observed in pDIS should be viewed as the quark spin defined in the light-cone formalism, which is different from the quark spin ($\Delta q_{QM}$) defined in the conventional quark model (or in the rest frame of the nucleon). These two quantities are connected by the relativistic effect of the Melosh-Wigner rotation~\cite{Melosh:1974cu,Wigner} due to the transversal motions of quarks inside the nucleon. The quantity $\Delta\Sigma/2=(\Delta u + \Delta d + \Delta s)/2$ represents, in a strict sense, the sum of the quark helicity in the light-cone frame rather than the vector sum of the spin carried by quarks and antiquarks in the proton rest frame. Hence, the best approach to analyze the experimental data is based on the framework of the light-cone field theory~\cite{Brodsky:1997de}, or equivalently the infinite momentum technique proposed by Weinberg~\cite{Weinberg:1966jm}. If we adopt the quantum field theory in the standard instant form, the spin of the proton should be the sum of the Melosh-Wigner rotated spin of the quarks rather than simply the sum of the spin of the individual constituent quarks. We would like to emphasize that although we shall pick a particular model to calculate the quark distribution functions, the relation between $\Delta q$ and $\Delta q_{QM}$ is general, since this effect is a fundamental kinematic relation that must be taken into account regardless of the dynamical details~\cite{Ma96,Bro2001}. However, not enough attention to this effect was paid in some previous analyses of the proton spin.

In this paper, we derive the polarized quark distribution functions $\Delta q(x)$ and the polarized structure functions $g_1(x)$ for the proton, the neutron, and the deuteron, respectively, in the framework of the chiral quark model without referring to polarized experimental data beforehand, and we show that the results can match the experimental data well through the fully relativistic light-cone treatment. The paper is organized as follows. In Sec.~\ref{section2}, we analyze the proton spin in the framework of the chiral quark model. In Sec.~\ref{section3}, we discuss the bare quark input in the chiral quark model and explain the important effect of the Melosh-Wigner rotation. Then we obtain the polarized quark distribution functions and the polarized structure functions for the proton, the neutron and the deuteron. In Sec.~\ref{section4}, we summarize the paper.

\section{Chiral Quark Model}\label{section2}

\subsection{The basic structure of the chiral quark model}

The chiral quark model~\cite{Weinberg:1978kz,Manohar:1983md} is widely recognized as an effective theory of QCD at the low-energy scale. It adopts a description of its important degrees of freedom in terms of quarks, gluons, and Goldstone bosons at momentum scales relating to hadron structure. In this model, the minor effects of the internal gluons are neglected, and the valence quarks contained in the nucleon fluctuate into quarks plus Goldstone bosons, which spontaneously break the chiral symmetry. This model is successful in explaining a number of problems, including the violation of the Gottfried sum rule from the aspect of the flavor asymmetry in the nucleon sea~\cite{Eichten:1991mt,Wakamatsu:1991tj,Song:2011fc}, the NuTeV anomaly resulting from the strange-antistrange asymmetry~\cite{Ding:2004dv}, and the isospin symmetry breaking between the proton and the neutron~\cite{Song:2010rq}. It is also proposed by Cheng and Li~\cite{Cheng:1994zn} that the proton spin problem can be accounted for by the quark splitting into a quark plus a Goldstone boson in the chiral quark model from an intuitive argument.

The effective Lagrangian of the interaction between Goldstone bosons and quarks in the leading order is
\begin{equation}
\mathcal{L}_{int}=-\frac{g_{A}}{f}\overline{\psi}\gamma^{\mu}\gamma_{5}\partial_{\mu}\Pi\psi,
\end{equation}
where $\Pi$ takes the form
\begin{equation}
\Pi\equiv\frac{1}{\sqrt{2}}\left(
\begin{array}{ccc}
  \frac{\pi^{0}}{\sqrt{2}}+\frac{\eta}{\sqrt{6}} & \pi^{+} & K^{+} \\
  \pi^{-} & -\frac{\pi^{0}}{\sqrt{2}}+\frac{\eta}{\sqrt{6}} & K^{0} \\
  K^{-} & \overline{K^{0}} & \frac{-2\eta}{\sqrt{6}} \\
\end{array}
\right).
\end{equation}
Based on the perturbative technique in the light-cone field theory, all particles are on mass shell and the factorization of the subprocess is applicable, so we can express the quark distributions inside a nucleon as a convolution of a constituent-quark distribution in a nucleon and the structure functions of a constituent quark. The Fock decompositions of constituent-quark wave functions are
\begin{eqnarray}
|U\rangle &=& \sqrt{Z}|u_{0}\rangle + a_{\pi}|d\pi^{+}\rangle + \frac{a_{\pi}}{\sqrt{2}}|u\pi^{0}\rangle + a_{K}|sK^{+}\rangle + \frac{a_{\eta}}{\sqrt{6}}|u\eta\rangle,\nonumber\\
|D\rangle &=& \sqrt{Z}|d_{0}\rangle + a_{\pi}|u\pi^{-}\rangle + \frac{a_{\pi}}{\sqrt{2}}|d\pi^{0}\rangle + a_{K}|sK^{0}\rangle + \frac{a_{\eta}}{\sqrt{6}}|d\eta\rangle,
\end{eqnarray}
where $Z=1-\frac{3}{2}\langle P_{\pi}\rangle-\langle P_{K}\rangle-\frac{1}{6}\langle P_{\eta}\rangle$ is the renormalization constant for the bare constituent, and $|a_{\alpha}|^{2}$ ($\alpha=\pi$, K, $\eta$) are the probabilities of finding Goldstone bosons in the dressed constituent-quark states $|U\rangle$ and $|D\rangle$. The fluctuation of a bare constituent quark into a Goldstone boson and a recoil bare constituent quark is given as~\cite{Suzuki:1997wv}
\begin{eqnarray}
q_{j}(x) &=& \int^{1}_{x}\frac{\textmd{d}y}{y}P_{i \rightarrow j\alpha}(y)q_{i}\left(\frac{x}{y}\right)\nonumber\\
&\equiv& P_{i \rightarrow j\alpha}\otimes \Delta q_{i 0} (x).
\end{eqnarray}
$P_{i \rightarrow j\alpha}(y)$ is the splitting function, which gives the probability of finding a constituent quark $j$ carrying the light-cone momentum fraction $y$ together with a spectator Goldstone boson $\alpha$,
\begin{eqnarray}
P_{i \rightarrow j\alpha}(y)&=&\frac{1}{32\pi^{2}}\left(\frac{g_{A}\left(m_i + m_j\right)}{f}\right)^{2} \int \mathrm{d}k^{2}_{\perp}\frac{(m_{j}-m_{i}y)^{2} + k^{2}_{\perp}}{y^{2}(1-y)[m_{i}^{2}-M^{2}_{j\alpha}]^{2}},
\end{eqnarray}
where $m_{i}$, $m_{j}$, and $m_{\alpha}$ are the masses of the $i$- and $j$-constituent quarks and the pseudoscalar meson $\alpha$ respectively. $M^{2}_{j\alpha}=\frac{m^{2}_{j}+k^{2}_{\perp}}{y} + \frac{m^{2}_{\alpha}+k^{2}_{\perp}}{1-y}$ is the square of the light-cone invariant mass of the final state.

In the effective field theory, it is necessary to introduce an ultraviolet cutoff to make the results finite. The conventional choice is to specify the momentum cutoff of the vertex as
\begin{equation}
g_{\mathrm{A}}\rightarrow g_{\mathrm{A}}^{\prime}\textmd{exp}\left[\frac{m^{2}_{i}-M^{2}_{j\alpha}}{4\Lambda^{2}}\right].
\end{equation}
In this paper, we adopt $g_{\mathrm{A}}^{\prime}=1$, following the large $N_{c}$ argument~\cite{Weinberg:1990xm}. The cutoff parameter $\Lambda$ can be determined by the experimental data of the Gottfried sum and the constituent-quark-mass inputs for the pion,
\begin{eqnarray}
S_{G}&=&\int^{1}_{0}\frac{\mathrm{d}x}{x}[F^{p}_{2}(x)-F^{n}_{2}(x)]\nonumber\\
&=&\frac{1}{3}\int_{0}^{1}\mathrm{d}x[u(x)+\overline{u}(x)-d(x)-\overline{d}(x)]\nonumber\\&
=&\frac{1}{3}(Z-\frac{1}{2}\left<P_{\pi}\right>+\left<P_{K}\right>+\frac{1}{6}\left<P_{\eta}\right>)\nonumber\\
&=&\frac{1}{3}(1-2\left<P_{\pi}\right>)\nonumber\\
&=& 0.235 \pm 0.026 ~~~~~\textmd{(experimental~data)},
\end{eqnarray}
where $\left<P_{\alpha}\right> = \int_0^1 P_{i \rightarrow j\alpha}(y) \mathrm{d}y$. We choose the parameters of masses as $m_u=m_d=330~$MeV, $m_s=480~$MeV, $m_{\pi}=140~$MeV, $m_K=495~$MeV, $m_{\eta}=495~$MeV, and therefore the cutoff $\Lambda = 1500~$MeV.

\subsection{Polarized quark distributions in the chiral quark model}

The polarized quark distributions are obtained by replacing the unpolarized splitting function $P_{i \rightarrow j\alpha}(y)$ with the polarized splitting function $\Delta P_{i \rightarrow j\alpha}(y) = P_{i\uparrow \rightarrow j\uparrow\alpha} - P_{i\downarrow \rightarrow j\downarrow\alpha}$. The polarized fluctuation process is expressed as
\begin{eqnarray}
\Delta q_{j}(x) &=& \int^{1}_{x}\frac{\textmd{d}y}{y}\Delta P_{i \rightarrow j\alpha}(y)\Delta q_{i 0}\left(\frac{x}{y}\right) \nonumber\\
&\equiv& \Delta P_{i \rightarrow j\alpha}\otimes \Delta q_{i 0} (x).
\end{eqnarray}
If $\Delta P_{i \rightarrow j\alpha}(y)$ is known, one can write down the polarized parton distribution functions
\begin{eqnarray}
\Delta u (x) &=& Z \Delta u_{0} (x)+ \frac{1}{2}\Delta P_{u \rightarrow u \pi^{0}}\otimes \Delta u_{0} (x) + \Delta P_{d \rightarrow u \pi^{-}}\otimes \Delta d_{0} (x) + \Delta P_{u \rightarrow u \eta}\otimes \Delta u_{0} (x), \nonumber\\
\Delta d (x) &=& Z \Delta d_{0} (x) + \frac{1}{2}\Delta P_{d \rightarrow d \pi^{0}}\otimes \Delta d_{0} (x) + \Delta P_{u \rightarrow d \pi^{+}}\otimes \Delta u_{0} (x) + \Delta P_{d \rightarrow d \eta}\otimes \Delta d_{0} (x), \nonumber\\
\Delta s (x) &=& \Delta P_{u \rightarrow s K^{+}}\otimes \Delta u_{0} (x) + \Delta P_{d \rightarrow s K^{0}}\otimes \Delta d_{0} (x), \nonumber\\
\Delta \overline{u} (x) &=& \Delta \overline{d} (x) = \Delta \overline{s} (x) = 0.
\end{eqnarray}
We should point out that in the chiral quark model, antiquarks are only produced in the process of the splitting of Goldstone bosons. Since Goldstone bosons are unpolarized, the polarization of antiquarks is zero, and this is compatible with the available experimental data~\cite{Airapetian:2007mh}.

\begin{figure*}
\begin{center}
\resizebox{0.95\textwidth}{!}
{\includegraphics{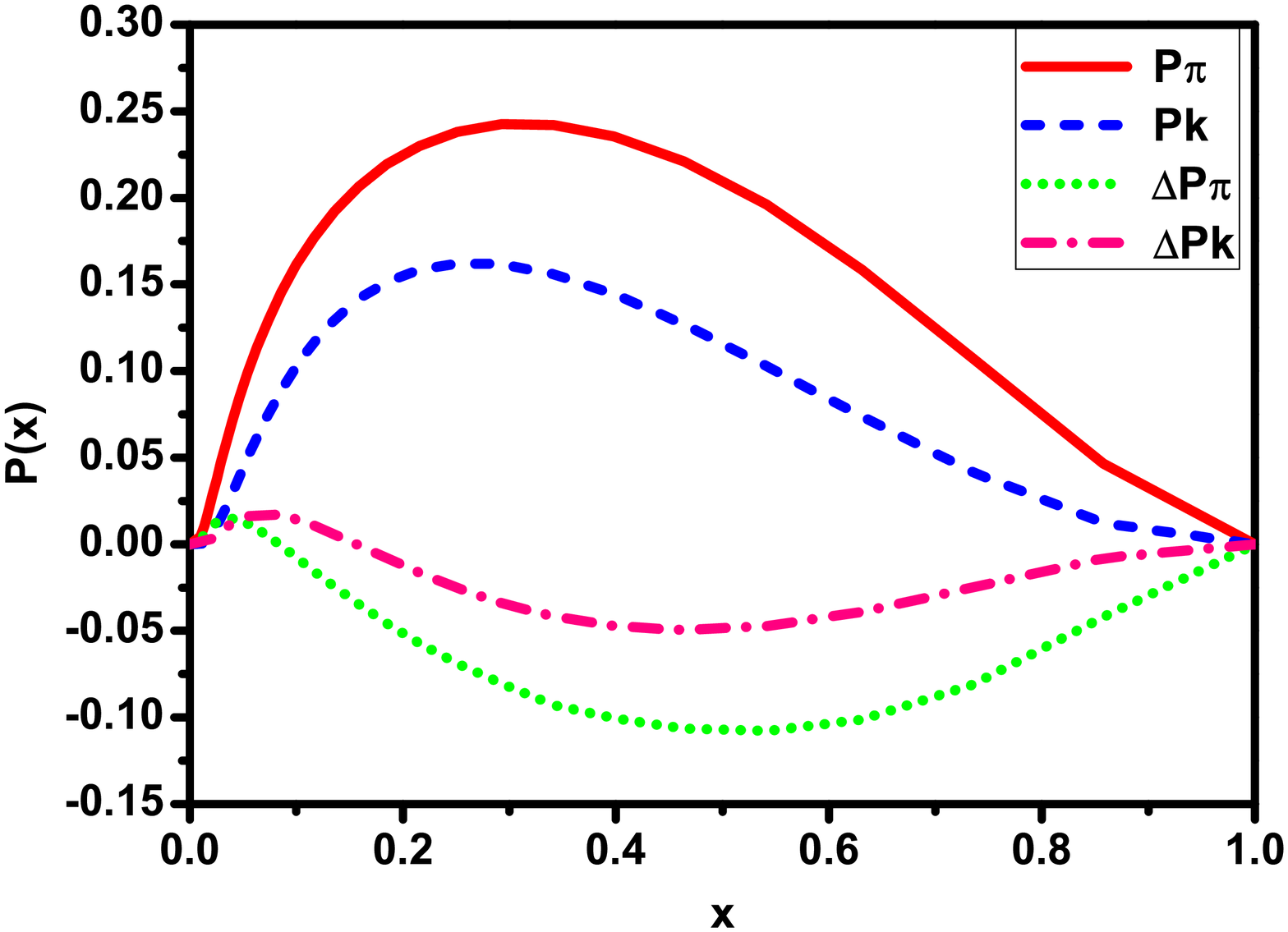}
\includegraphics{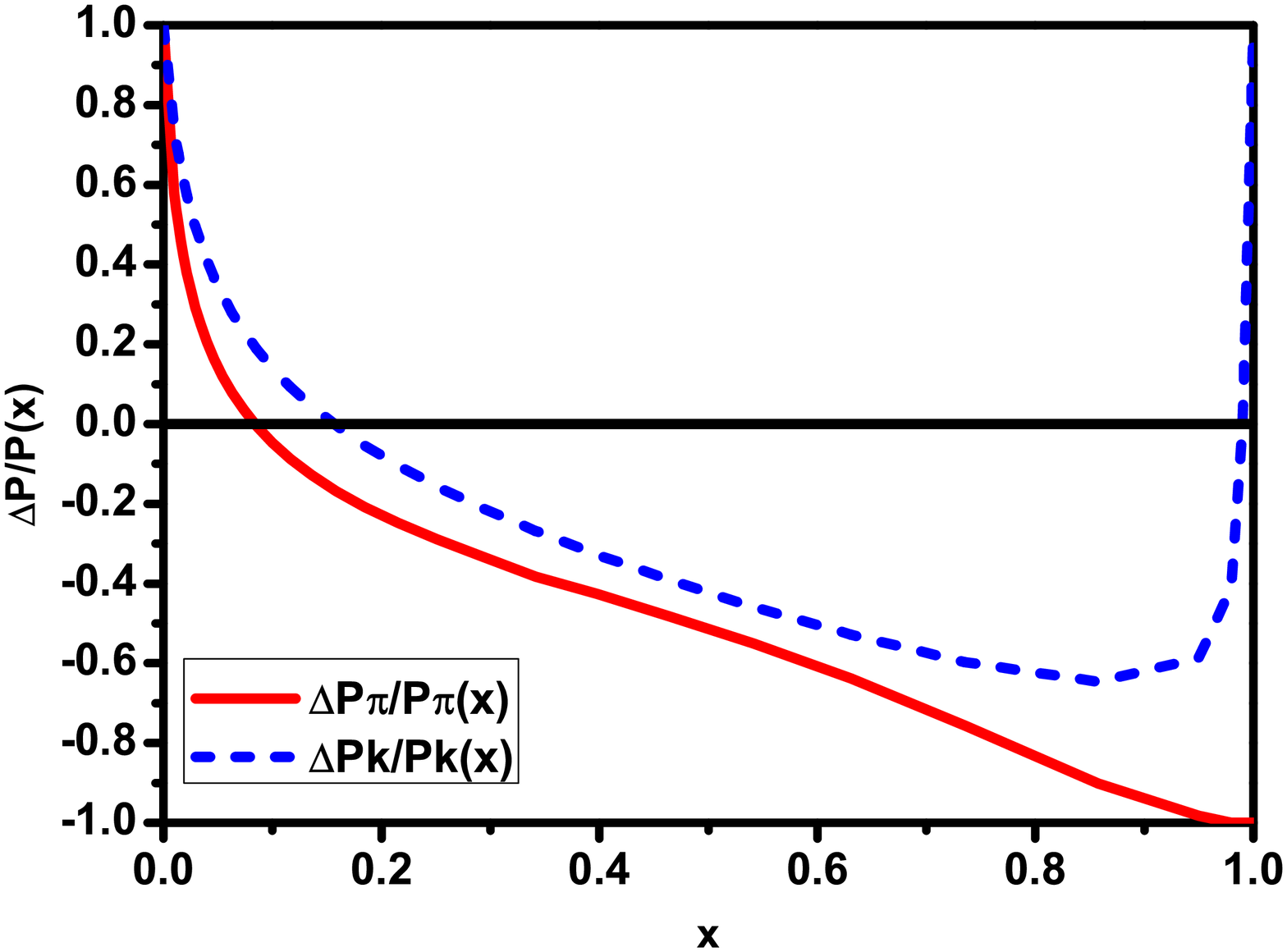}}
\caption{\small The behavior of the unpolarized and polarized splitting functions $P$ and $\Delta P$ versus $x$. }\label{p}
\end{center}
\end{figure*}

In most models including the chiral quark model, the possibility of ``helicity reverse" is correlated with the matrix element
\begin{equation}
\frac{\overline{u}\left(k, \uparrow (\downarrow)\right)}{\sqrt{k^{+}}} \gamma_{5} \frac{u\left(p, \uparrow (\downarrow)\right)}{\sqrt{p^{+}}},
\end{equation}
which describes the change of the helicity of a fermion after interacting with a pseudoscalar particle. It is often taken as an assumption~\cite{Cheng:1994zn} that the helicity of the fermion is totally reversed, and therefore $\Delta P_{i \rightarrow j\alpha}(y) = - P_{i \rightarrow j\alpha}(y)$ in the explanation of the proton spin problem. However, as stressed in the introduction, we should adopt the light-front formula~\cite{Brodsky:1997de} to analyze the spin structure of a composed system carefully. In this approach we can obtain
\begin{eqnarray}
\frac{\overline{u}\left(k, m, \uparrow\right)}{\sqrt{k^{+}}}
\gamma_{5} \frac{u\left(p, M, \uparrow\right)}{\sqrt{p^{+}}} =
\frac{2}{2 k^{+} p^{+}}\left[- k^{+} p^{+} + m M - \left(k^{1}-i
k^{2}\right)\left(p^{1} + i p^{2}\right)\right],\\
\frac{\overline{u}\left(k, m, \uparrow\right)}{\sqrt{k^{+}}}
\gamma_{5} \frac{u\left(p, M, \downarrow\right)}{\sqrt{p^{+}}} = -
\frac{2}{2 k^{+} p^{+}}\left[m \left(p^{1} - i p^{2}\right) + M \left(k^{1} - i k^{2}\right)\right],\\
\frac{\overline{u}\left(k, m, \downarrow\right)}{\sqrt{k^{+}}}
\gamma_{5} \frac{u\left(p, M, \downarrow\right)}{\sqrt{p^{+}}} =
\frac{2}{2 k^{+} p^{+}}\left[ k^{+} p^{+} - m M + \left(k^{1}+i
k^{2}\right)\left(p^{1} - i p^{2}\right)\right],\\
\frac{\overline{u}\left(k, m, \downarrow\right)}{\sqrt{k^{+}}}
\gamma_{5} \frac{u\left(p, M, \uparrow\right)}{\sqrt{p^{+}}} = -
\frac{2}{2 k^{+} p^{+}}\left[m \left(p^{1} + i p^{2}\right) + M
\left(k^{1} + i k^{2}\right) \right].
\end{eqnarray}
Furthermore, the following kinematics is generally associated with the vertex
\begin{eqnarray}
p = \left( p^{+}, \frac{m_i^2}{p^{+}}, \overrightarrow{0_{\perp}} \right),~~~~~ k = \left( y p^{+}, \frac{m_j^2+\overrightarrow{k_{\perp}}^2}{y p^{+}}, \overrightarrow{k_{\perp}}
\right).
\end{eqnarray}
We obtain the form of the helicity reverse function $\Delta P_R$,
\begin{equation}
\Delta P_R \equiv P\left(\uparrow\Rightarrow\uparrow -
\uparrow\Rightarrow\downarrow\right) = \frac{\left(-y p^{+ 2}+m_i
m_j\right)^2 - m_i^2 \overrightarrow{k_{\perp}}^2}{\left(-y p^{+ 2}+m_i
m_j\right)^2 + m_i^2 \overrightarrow{k_{\perp}}^2},
\end{equation}
which is not $-1$. Accordingly, in the chiral quark model, the polarized splitting function $\Delta P_{i \rightarrow j\alpha}(y)$ takes the form
\begin{eqnarray}
\Delta P_{i \rightarrow j\alpha}(y) &=& P_{i \rightarrow j\alpha}(y) \Delta P_R \nonumber\\
&=&\frac{1}{32\pi^{2}}\left(\frac{g_{A}\left(m_i + m_j\right)}{f}\right)^{2} \int \textmd{d}k^{2}_{\perp}\frac{(m_{j}-m_{i}y)^{2} - k^{2}_{\perp}}{y^{2}(1-y)[m_{i}^{2}-M^{2}_{j\alpha}]^{2}}.
\end{eqnarray}
The behaviors of $P$ and $\Delta P$ are displayed in Fig.~\ref{p}.

It is important to distinguish two different types of quark and gluon contributions to the nucleon sea measured in experiments. One is the perturbative ``extrinsic" sea, which is generated from the QCD hard bremsstrahlung and the gluon-splitting process, and this type of sea is associated with the internal composition of gluons rather than the proton itself. The other one is the nonperturbative ``intrinsic" sea, which is directly connected to the valence quarks of the nucleon. The quark distributions measured at certain $Q^{2}$ include not only the intrinsic sea but also the extrinsic sea~\cite{Brodsky:1996hc}. However, in the chiral quark model, only the intrinsic sea is calculated while the extrinsic sea is left out. Therefore, it is generally necessary to add the extrinsic sea before comparing results with experiments when we consider the unpolarized quark distributions~\cite{Song:2011fc}. Nevertheless, when we consider the polarized quark distributions, we assume that the perturbative extrinsic sea is totally unpolarized, since the correlative processes give equal contributions to both types of polarization. Therefore we do not need to add the perturbative extrinsic sea.

\section{The Bare Quark Input and the Melosh-Wigner Rotation}\label{section3}

\subsection{The bare quark input}

We need the input of polarized bare quark distributions $\Delta u_{0}(x)$ and $\Delta d_{0}(x)$ in the calculation. However, since these distributions are not measurable in experiments, there is no parametrization available. It is also not proper to use polarized parton distribution parametrization of valance quark as the input directly, since these two kinds of distributions are physically different. In this paper, we try to determine them from the unpolarized quark distributions, since the chiral quark model should give a unified description of the unpolarized and the polarized quark distributions. Therefore, we actually use the chiral quark model as a theoretical tool to connect the unpolarized and the polarized quark distributions, and we never refer to polarized experimental data beforehand.

We assume that the unpolarized bare quark distributions $u_{0}(x)$ and $d_{0}(x)$ take the form
\begin{eqnarray}
u_{0} (x) = \frac{2 x^{a_{u}}(1-x)^{b_{u}}}{\textmd{B}\left[a_{u}+1,b_{u}+1\right]},\nonumber\\
d_{0} (x) = \frac{x^{a_{d}}(1-x)^{b_{d}}}{\textmd{B}\left[a_{d}+1,b_{d}+1\right]}.
\end{eqnarray}
where $B[i,j]$ is the Euler beta function. It is observed that $q(x)\rightarrow Z q_{0}(x)$ when $x \rightarrow 1$ in the chiral quark model, so we can determine $b_{u}$ and $b_{d}$ from the asymptotic relation given by the unpolarized parametrization. Afterward, we determine the value of $a_{u}$ and $a_{d}$ by requiring that the maximum points of functions $x q(x)~(q = u, d)$ should be at the same values of $x$ as those in the parametrization. It is found that the approximate values for these parameters are $a_{u} = 0.5$, $a_{d} = 1$, $b_{u} = 3$, $b_{u} = 5$ based on the Martin-Roberts-Stirling-Thorne parametrization~\cite{Martin:2009iq} at fixed $Q^2 = 2.5$~GeV$^2$. Then we can adopt the naive SU(6) quark model to get the spin structure of bare quarks in the proton rest frame,
\begin{equation}
\Delta u_{0}^{QM}(x) = \frac{2}{3} u_{0} (x), ~~~~~\Delta
d_{0}^{QM}(x) = -\frac{1}{3} d_{0} (x).
\end{equation}
Before obtaining the result in the light-cone frame, we need to analyze the effect of the Melosh-Wigner rotation carefully.

\subsection{The Melosh-Wigner rotation}

The Melosh-Wigner rotation~\cite{Melosh:1974cu,Wigner} is a natural result of relativistic kinematics, and was shown to play an essential role in the analysis of the spin structure~\cite{Ma:1991xq,Ma:1992sj}. The key point lies in the fact that the vector sum of the constituent spin for a composite system is not Lorentz invariant from the relativistic viewpoint. The quark helicity distribution function $\Delta q$ measured in pDIS is defined by the matrix element
\begin{equation}
\Delta q = \langle p^{\uparrow} | \overline{q} \gamma^{+} \gamma_{5} q | p^{\uparrow} \rangle.
\end{equation}
If we express the quark wave functions in terms of light-cone Dirac spinors~\cite{Brodsky:1997de}, we will obtain
\begin{equation}
\Delta q (x)=q^{\uparrow}(x)-q^{\downarrow}(x),
\end{equation}
where $q^{\uparrow}(x)$ and $q^{\downarrow}(x)$ are probabilities of finding a quark of flavor $q$ with fraction $x$ of the proton longitudinal momentum and with polarization parallel and antiparallel to the direction of the proton spin in the proton infinite momentum frame (i.e., the quark spin states in the light-cone frame). However, if we express the quark wave functions in terms of conventional instant form Dirac spinors (i.e., the quark spin states in the proton rest frame), we will get that
\begin{eqnarray}
\Delta q (x) &=& \int \mathrm{d}^{2}\overrightarrow{k_{\perp}} W_D \left(x, \overrightarrow{k_{\perp}}\right)\left[q_{s_z = \frac{1}{2}}\left(x, \overrightarrow{k_{\perp}}\right)-q_{s_z = -\frac{1}{2}}\left(x, \overrightarrow{k_{\perp}}\right)\right]\nonumber\\
&=& \int \mathrm{d}^{2}\overrightarrow{k_{\perp}} W_D \left(x, \overrightarrow{k_{\perp}}\right) \Delta q_{QM} \left(x, \overrightarrow{k_{\perp}}\right),
\end{eqnarray}
with
\begin{equation}
W_D \left( x, \overrightarrow{k_{\perp}}\right) = \frac{\left(k^{+}+m\right)^2-\overrightarrow{k_{\perp}}^2}{\left(k^{+}+m\right)^2+\overrightarrow{k_{\perp}}^2}
\end{equation}
being the contribution from the relativistic effect due to quark transversal motions, $q_{s_z = \frac{1}{2}}\left(x, \overrightarrow{k_{\perp}}\right)$ and $q_{s_z = -\frac{1}{2}}\left(x, \overrightarrow{k_{\perp}}\right)$ being probabilities of finding, in the proton rest frame, a quark or antiquark of flavor $q$ with rest mass $m$ and momentum $p_{\mu}$ and with spin parallel and antiparallel to the rest proton spin respectively, and $k^{+} = x {\cal M}$ with ${\cal M}^2 = \sum_{i}(m^2_i+ \overrightarrow{k_{i \perp}}^2) / {x_i}$.

\begin{figure}
\begin{center}
\includegraphics[width=0.95\textwidth]{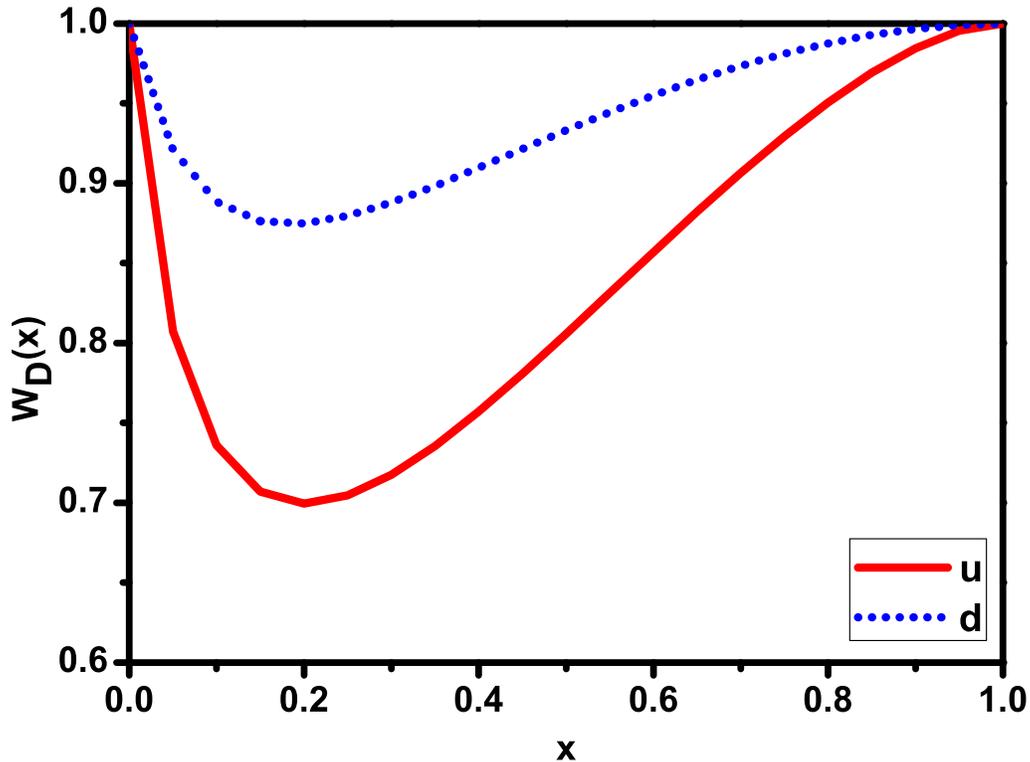}
\caption{\small The Melosh-Wigner rotation factor for $u$ quark and $d$ quark. The solid line and the dotted line stand for the results of $u$ quark and $d$ quark respectively. The parameters are chosen as $m_q = 330~$MeV, $m_{Du}=600~$MeV, $m_{Dd}=900~$MeV, $\alpha_{u}=330~$MeV, and $\alpha_{d}=200~$MeV.}\label{mw}
\end{center}
\end{figure}

It is straightforward to express the quark spin in the light-cone frame with the effect of the Melosh-Wigner rotation as
\begin{eqnarray}
\Delta u_{0}(x) = \Delta u_{0}^{QM}(x) W_u (x) = \frac{2}{3} u_{0} (x) W_u (x),\nonumber\\
\Delta d_{0}(x) = \Delta d_{0}^{QM}(x) W_d (x) = -\frac{1}{3} d_{0} (x) W_d (x),
\end{eqnarray}
where $W_u(x)$ and $W_d(x)$ are Melosh-Wigner rotation factors for the $u$ quark and the $d$ quark, respectively. In this paper, we adopt $m_q = 330~$MeV, $m_{Du}=600~$MeV, $m_{Dd}=900~$MeV, $\alpha_{u}=330~$MeV, and $\alpha_{d}=200~$MeV as inputs for the Melosh-Wigner rotation factor. The Melosh-Wigner rotation factors for the $u$ quark and the $d$ quark are displayed in Fig.~\ref{mw}. The difference of inputs for the Melosh-Wigner rotation between $u$ quark and $d$ quark results from the difference between the $u$ quark and the $d$ quark in the wave function~(\ref{wavefunction}). When the $u$ quark is probed, the spectator quarks have both the scalar ($S=0$) and the vector ($S=1$) components, but when the $d$ quark is probed, the spectator quarks only have the vector ($S=1$) component. Therefore, the $u$ quark has a more significant effect than the $d$ quark under Melosh-Wigner rotation.

\subsection{The polarized quark distribution functions}

\begin{figure}
\begin{center}
\includegraphics[width=0.95\textwidth]{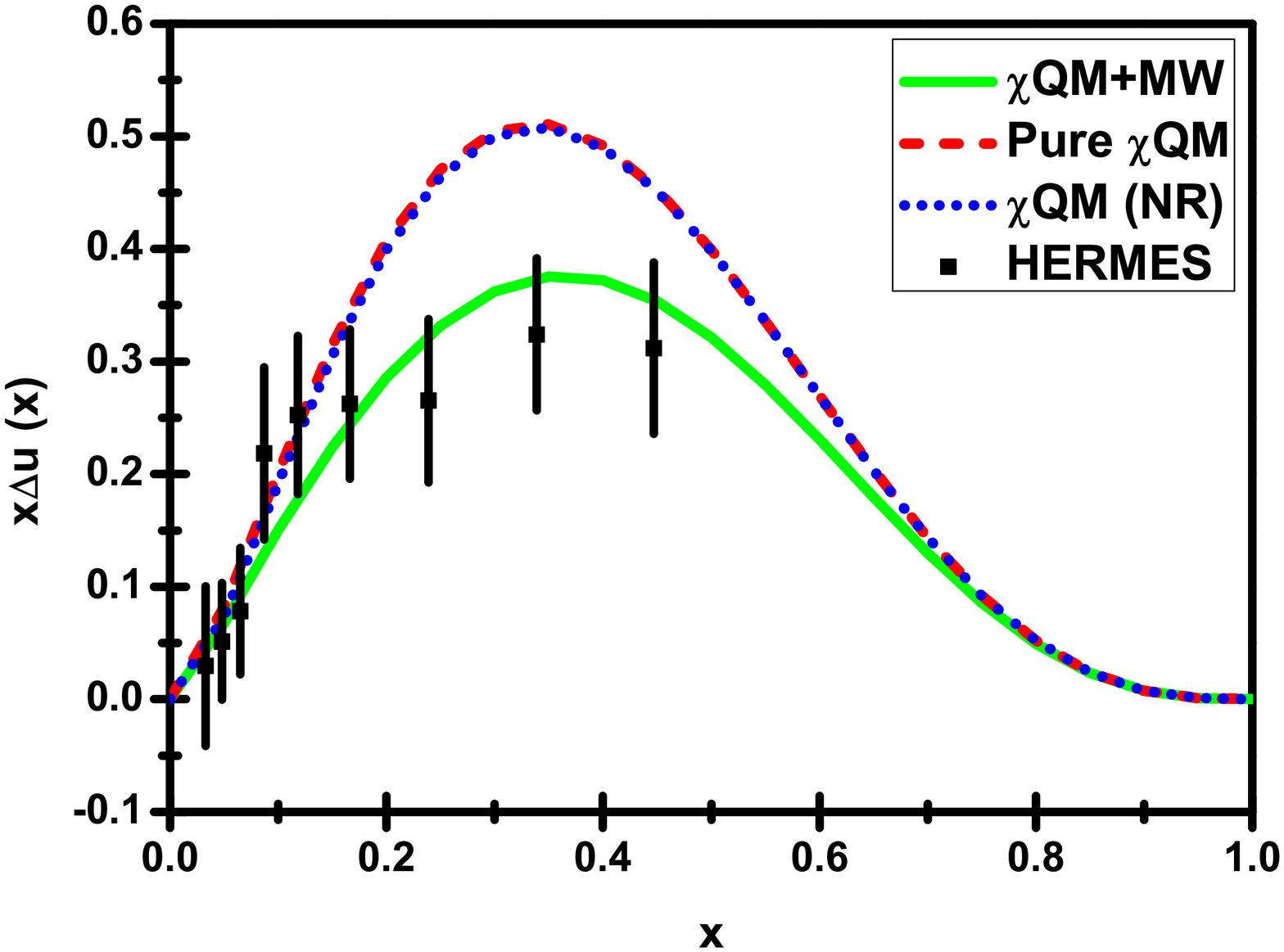}
\caption{\small The polarized quark distribution function for $u$
quark $x \Delta u(x)$ versus $x$. The solid line is the result
from the chiral quark model with Melosh-Wigner rotation taken into
account. The dashed line is the result from the pure chiral
quark model without taking Melosh-Wigner rotation into account. The
dotted line is the result from the chiral quark model using the
nonrelativistic vertex. The experimental data of HERMES are from
Ref.~\cite{Airapetian:2007mh} and evaluated at a common $Q^2 = 2.5~$GeV$^2$.}\label{delu}
\end{center}
\end{figure}

\begin{figure}
\begin{center}
\includegraphics[width=0.95\textwidth]{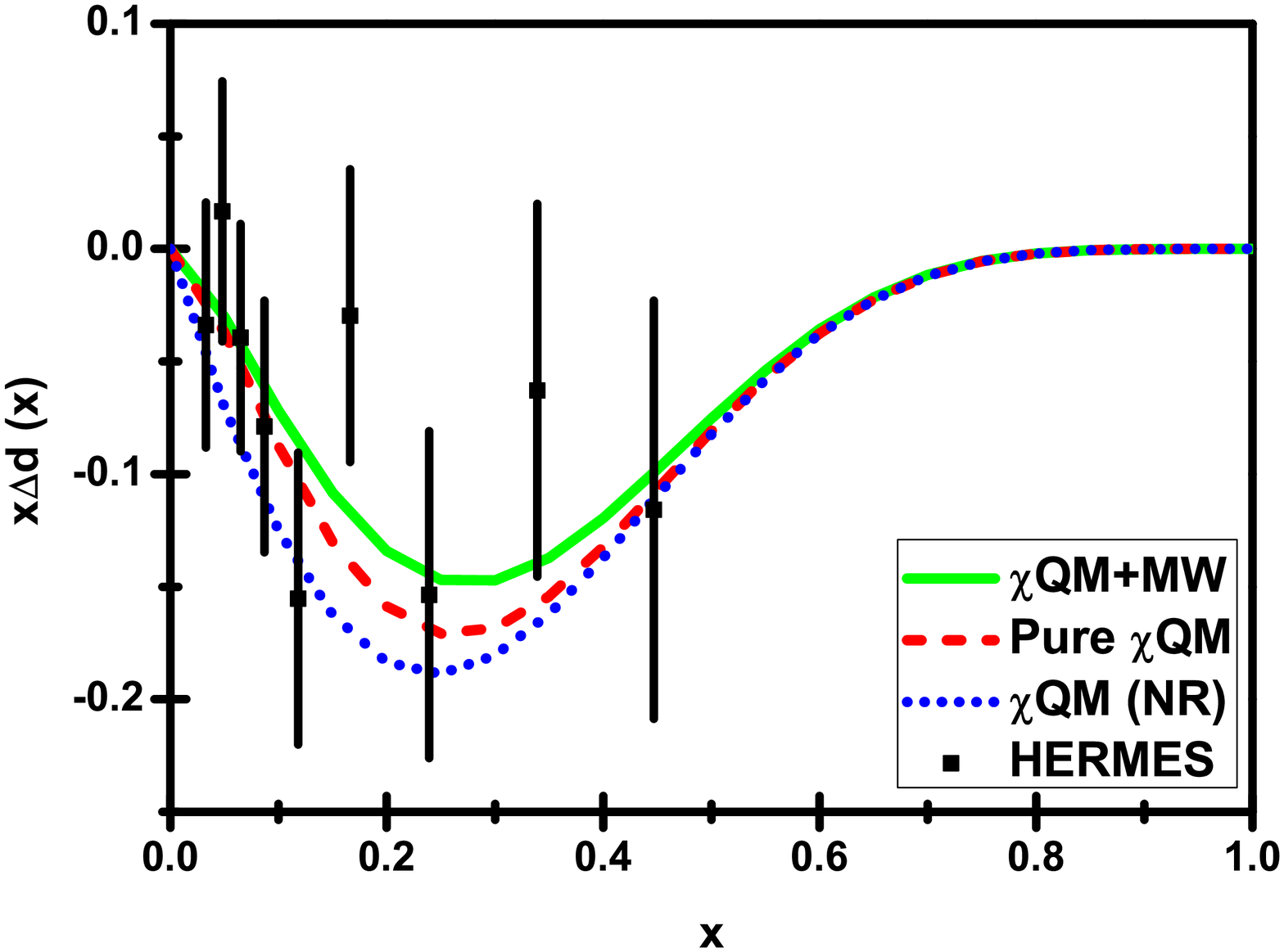}
\caption{\small The polarized quark distribution function for $d$
quark $x \Delta d(x)$ versus $x$. The solid line is the result
from the chiral quark model with Melosh-Wigner rotation taken into
account. The dashed line is the result from the pure chiral
quark model without taking Melosh-Wigner rotation into account. The
dotted line is the result from the chiral quark model using the
nonrelativistic vertex. The experimental data of HERMES are from
Ref.~\cite{Airapetian:2007mh} and evaluated at a common $Q^2 = 2.5~$GeV$^2$.}\label{deld}
\end{center}
\end{figure}

\begin{figure}
\begin{center}
\includegraphics[width=0.95\textwidth]{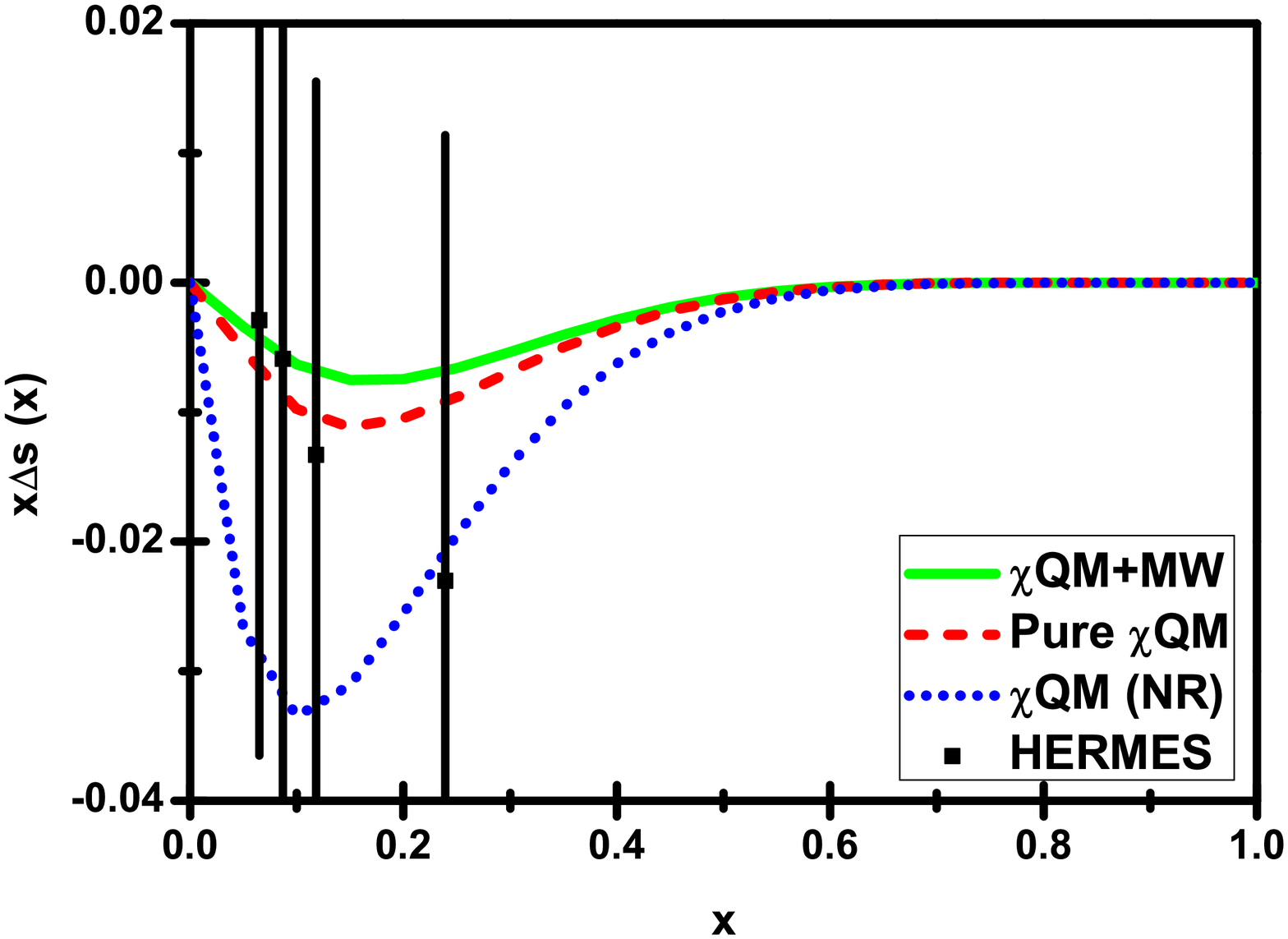}
\caption{\small The polarized quark distribution function for $s$
quark $x \Delta s(x)$ versus $x$. The solid line is the result
from the chiral quark model with Melosh-Wigner rotation taken into
account. The dashed line is the result from the pure chiral
quark model without taking Melosh-Wigner rotation into account. The
dotted line is the result from the chiral quark model using the
nonrelativistic vertex. The experimental data of HERMES are from
Ref.~\cite{Airapetian:2007mh} and evaluated at a common $Q^2 = 2.5~$GeV$^2$.}\label{dels}
\end{center}
\end{figure}

With all the above considerations taken into account, we can get the polarized quark distribution functions $\Delta u(x)$, $\Delta d(x)$, and $\Delta s(x)$, and the results are displayed in Figs.~\ref{delu},~\ref{deld},~\ref{dels}. It should be pointed out that the experimental data have been evaluated to a common $Q^2 = 2.5~$GeV$^2$ in order to compare with the theoretical prediction, which is also conducted at the same $Q^2$. It can be found that our results match with the experimental data well.

\begin{figure*}
\begin{center}
{\includegraphics[width=0.45\textwidth]{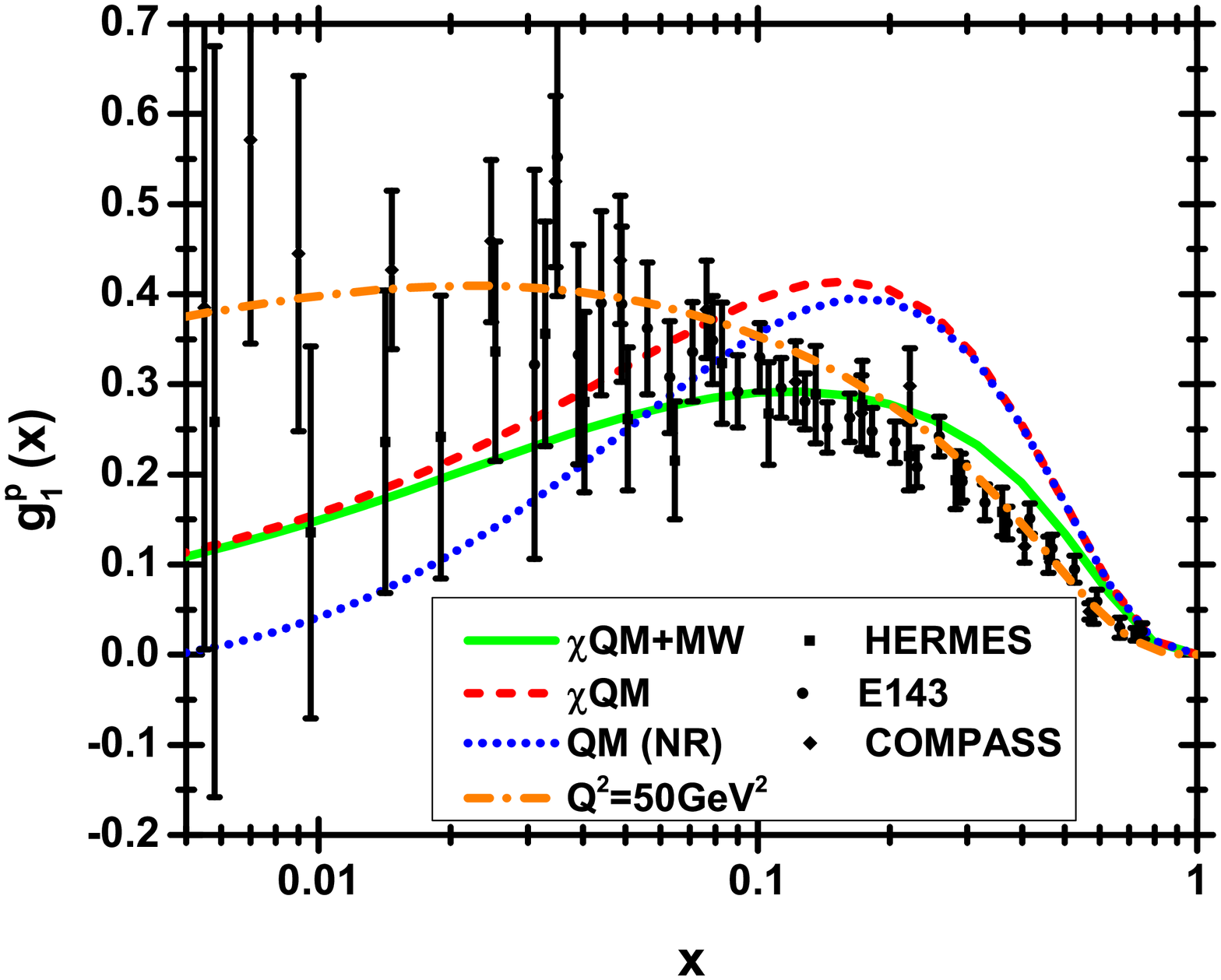}
\includegraphics[width=0.45\textwidth]{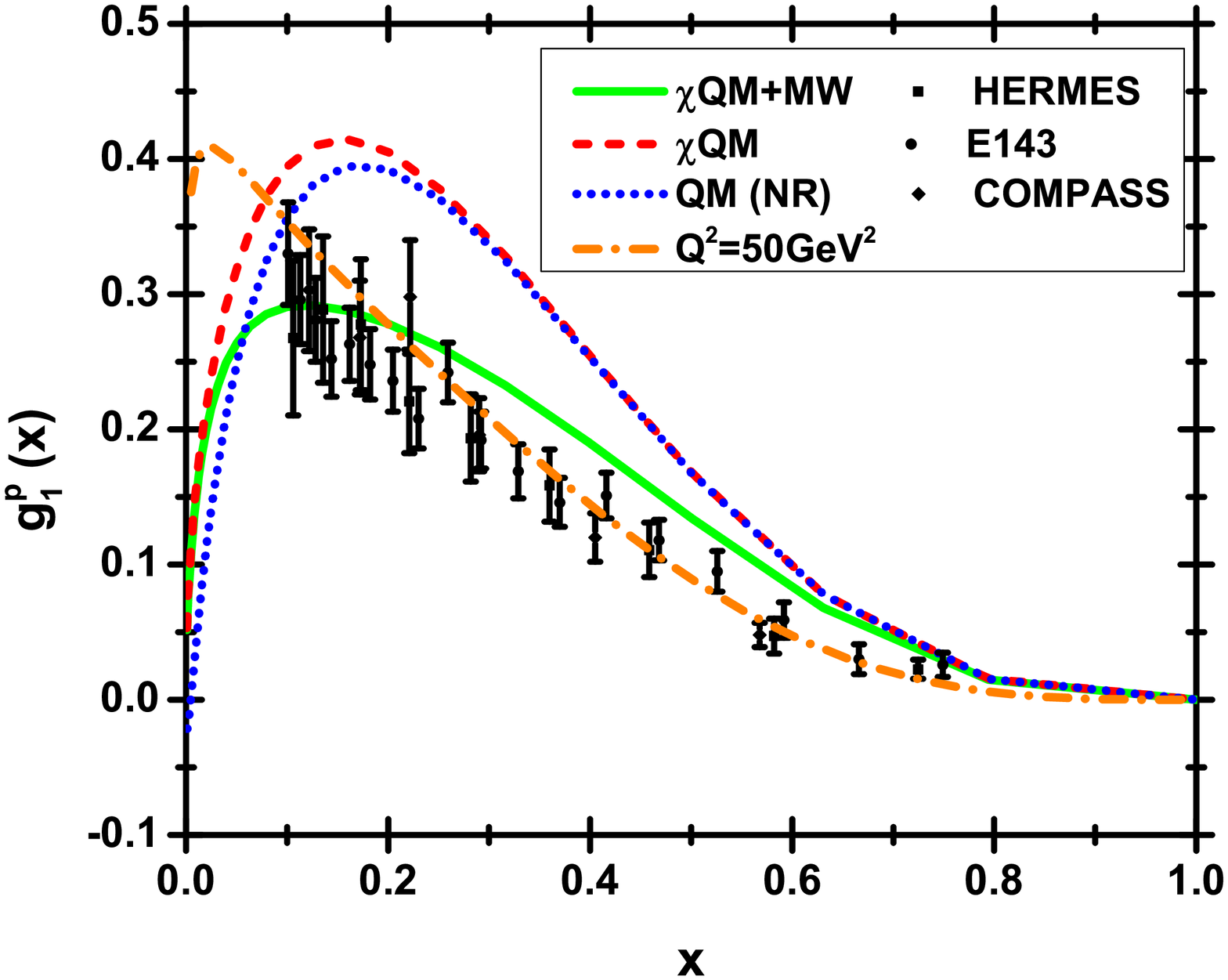}}
\caption{\small The polarized structure function for the proton
$g_1^p$ versus $x$. The solid line is the result from the chiral
quark model with Melosh-Wigner rotation taken into account. The
dashed line is the result from the pure chiral quark model without
taking Melosh-Wigner rotation into account. The dotted line is the
result from the chiral quark model using the nonrelativistic
vertex. The dash-dotted line is the result after evolution to $Q^2 =
50~$GeV$^2$. The experimental data of HERMES, E143, and COMPASS are
from Refs.~\cite{Airapetian:2007mh,Abe:1998wq,Ageev:2005gh,Alekseev:2010hc},
respectively. The left and right figures differ in the $x$-axis
type for the purpose of showing the small-$x$ (left figure) and
large-$x$ (right figure) behaviors. We only include data with $x>0.1$ in the right figure.}\label{g1p}
\end{center}
\end{figure*}

\begin{figure*}
\begin{center}
{\includegraphics[width=0.45\textwidth]{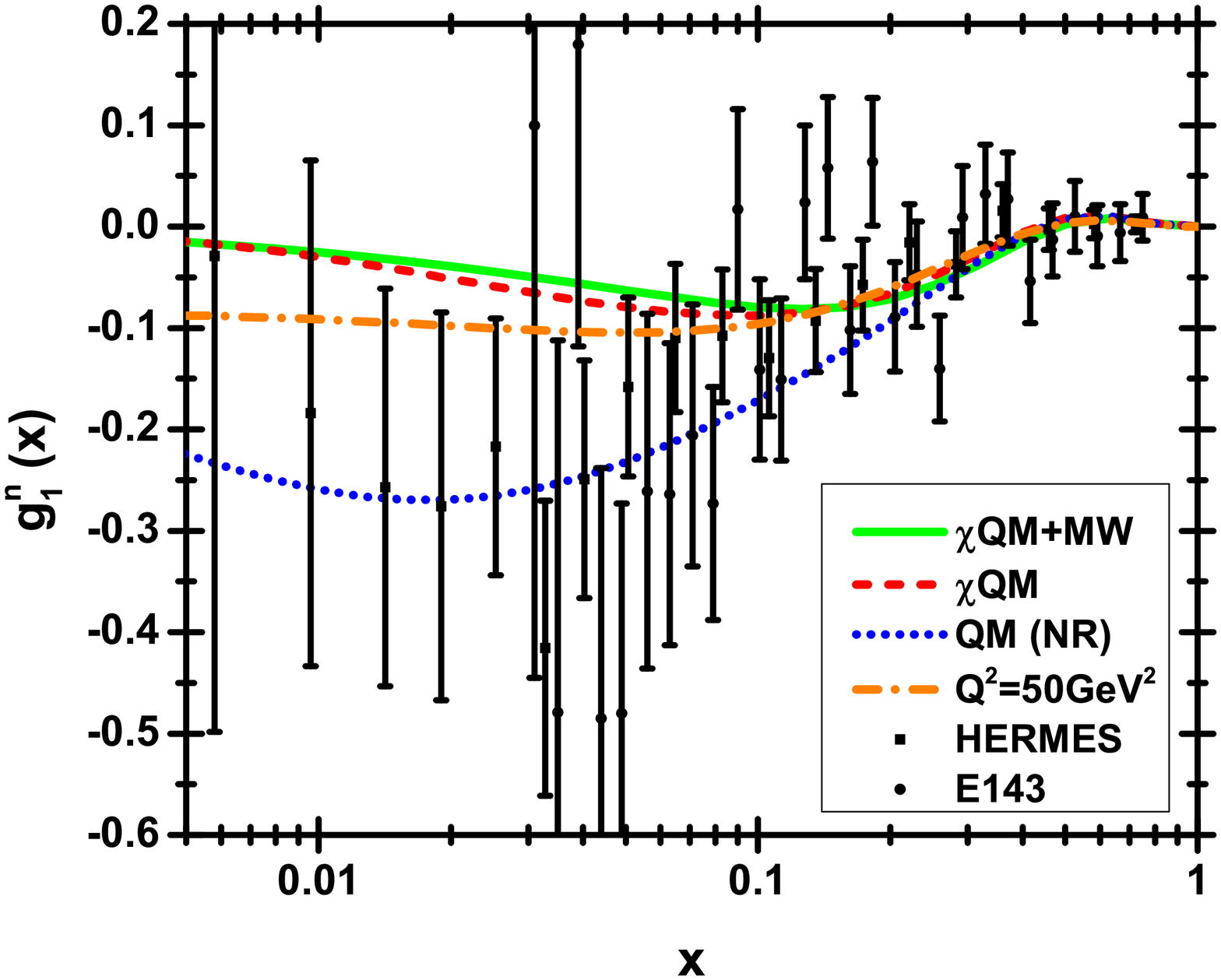}
\includegraphics[width=0.45\textwidth]{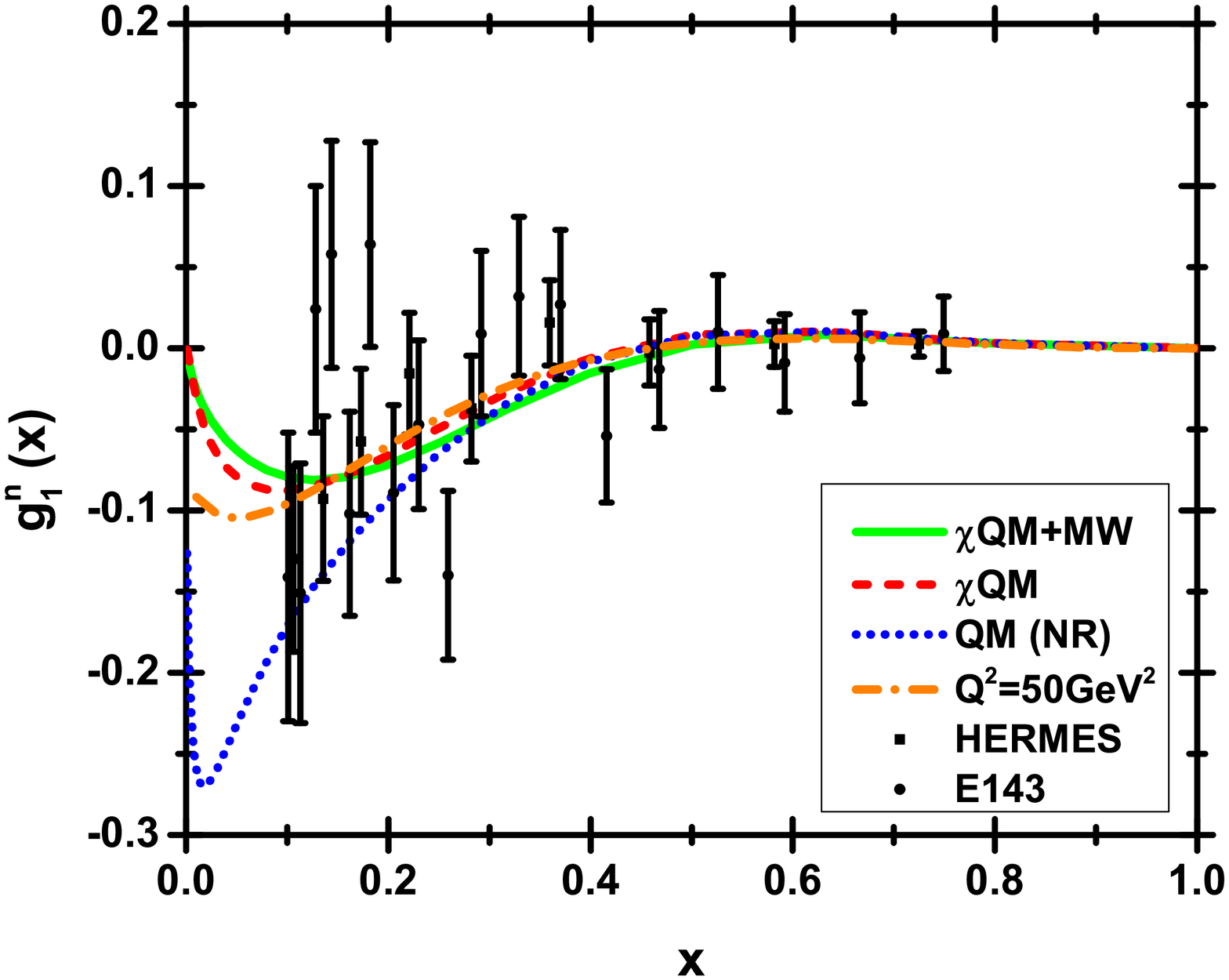}}
\caption{\small The polarized structure function for the neutron
$g_1^n$ versus $x$. The solid line is the result from the chiral
quark model with Melosh-Wigner rotation taken into account. The
dashed line is the result from the pure chiral quark model without
taking Melosh-Wigner rotation into account. The dotted line is the
result from the chiral quark model using the nonrelativistic
vertex. The dash-dotted line is the result after evolution to $Q^2 =
50~$GeV$^2$. The experimental data of HERMES and E143 are from
Refs.~\cite{Airapetian:2007mh,Abe:1998wq}, respectively. The left and right figures differ in the $x$-axis
type for the purpose of showing the small-$x$ (left figure) and large-$x$ (right figure) behaviors.  We only include data with $x>0.1$ in the right figure.}\label{g1n}
\end{center}
\end{figure*}

\begin{figure*}
\begin{center}
{\includegraphics[width=0.45\textwidth]{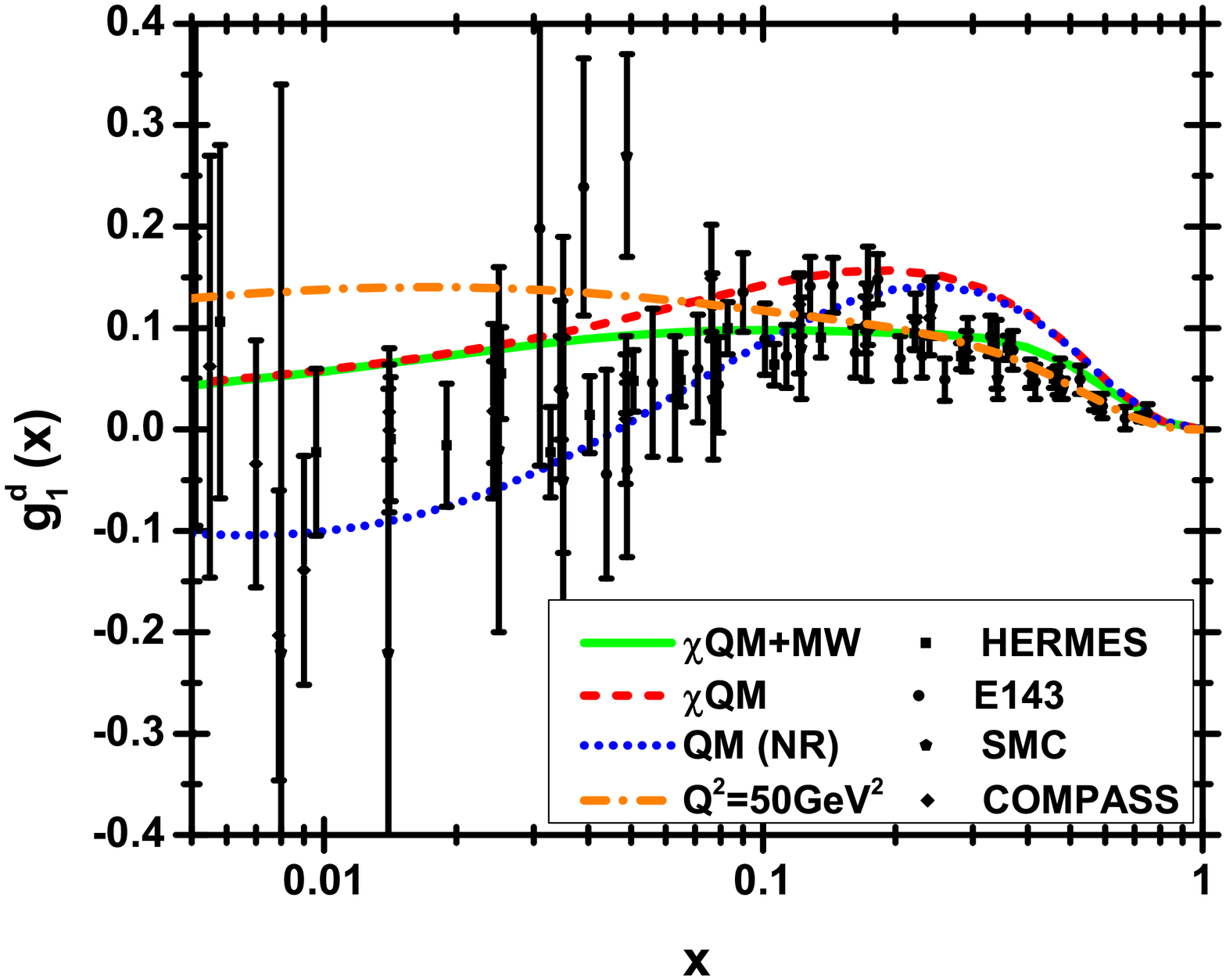}
\includegraphics[width=0.45\textwidth]{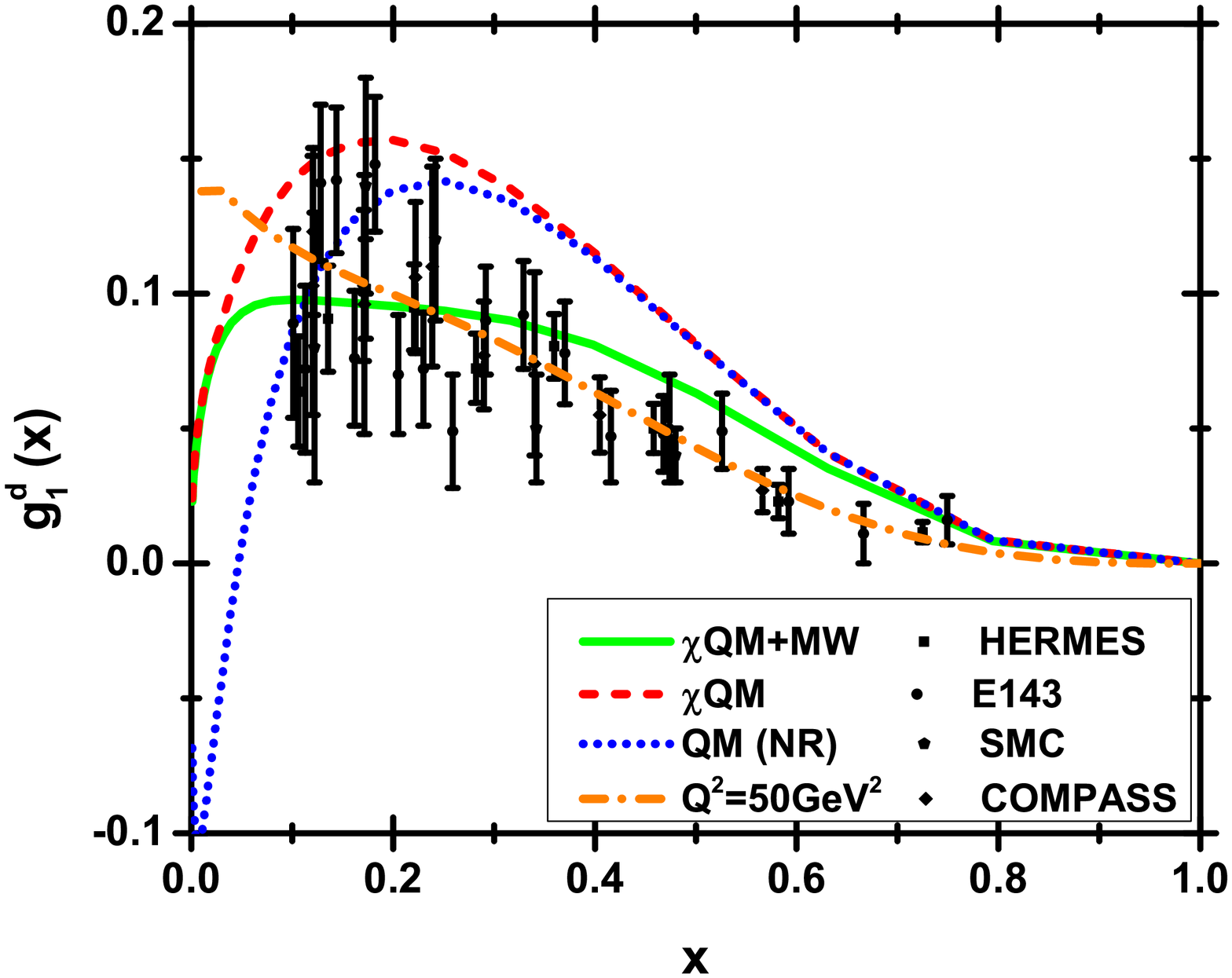}}
\caption{\small The polarized structure function for the deuteron
$g_1^d$ versus $x$. The solid line is the result from the chiral
quark model with Melosh-Wigner rotation taken into account. The
dashed line is the result from the pure chiral quark model without
taking Melosh-Wigner rotation into account. The dotted line is the
result from the chiral quark model using the nonrelativistic
vertex. The dash-dotted line is the result after evolution to $Q^2 =
50~$GeV$^2$. The experimental data of HERMES, E143, SMC, and COMPASS
are from Refs.~\cite{Airapetian:2007mh,Abe:1998wq,Adeva:1998vv,Ageev:2005gh,Alekseev:2010hc},
respectively. The left and right figures differ in the $x$-axis
type for the purpose of showing the small-$x$ (left figure) and
large-$x$ (right figure) behaviors.  We only include data with $x>0.1$ in the right figure.}\label{g1d}
\end{center}
\end{figure*}

Furthermore, we display the polarized structure functions $g_1^N, (N = p, n, d)$ in Figs.~\ref{g1p},~\ref{g1n},~\ref{g1d}. It is shown that when we use the chiral quark model with the effect of the Melosh-Wigner rotation taken into consideration, the behaviors of these functions can match the experimental data well. In contrast, when the Melosh-Wigner rotation is not included, the description of the behavior in large $x$ region is not satisfying. Hence, the
Melosh-Wigner rotation can provide an important mechanism to depress the polarization in large $x$ region.

To make our analysis more comprehensive and persuasive, we also
evaluate the polarized quark distribution functions to a much higher
energy scale $Q^2 = 50~$GeV$^2$ with the Higher Order Perturbative
Parton Evolution Toolkit (HOPPET)~\cite{Salam:2008qg}, and show the
behavior of $g_1^N$ at $Q^2 = 50~$GeV$^2$ in
Figs.~\ref{g1p},~\ref{g1n},~\ref{g1d} respectively. We find that the
$Q^2$-evoluted results can match better with the experimental data
at large $x$, in which region $Q^2$ is much larger than
$2.5~$GeV$^2$, and also the results are insensitive to different
inputs of gluon contributions. However, the experimental precision
is still not good enough to distinguish the effect of
$Q^2$ evolution and we expect more experimental investigations
concerning the QCD evolution behaviors of spin related quantities of
the nucleon in the future.


\section{Summary}\label{section4}

In this paper, we discuss the spin structure of the proton in the
relativistic light-cone formulae. We stress that the quark helicity
measured in pDIS is actually the quark spin defined in the
light-cone formalism and is different from the quark spin defined
in the quark model (or rest frame of the nucleon), and these two
quantities are related by the relativistic kinematical effect of the
Melosh-Wigner rotation. This effect must be taken into account
regardless of the dynamical details. We adopt the chiral quark model
to calculate the polarized quark distribution functions $\Delta
q(x)$ and the polarized structure functions $g_1$. We show that the
results can match the experimental data well. However, we would like
to emphasize that the chiral quark model may possibly be replaced
by some other kinds of dynamics, but the relativistic kinematics of
the Melosh-Wigner rotation is prerequisite. We suggest that more
precise experiments should be carried out to enable more accurate
determination of polarized quark distributions to confront with
theoretical calculations.

\section*{Acknowledgments}

We thank Junwu Huang and Tianbo Liu for their help in obtaining the QCD
evolution results. This work is supported by the National Natural
Science Foundation of China (Grants No.~10721063, No.~10975003,
No.~11035003, and No.~11120101004), and the National Fund for Fostering Talents
of Basic Science (Grant No.~J1030310).

\end{document}